\documentclass[%
 aip,
 amsmath,amssymb,
 reprint,%
]{revtex4-1}

\usepackage{graphicx}
\usepackage{dcolumn}
\usepackage{bm}
\usepackage[utf8]{inputenc}
\usepackage[T1]{fontenc}
\usepackage{mathptmx}

\begin{document}

\preprint{AIP/123-QED}

\title[Hwang \textit{et al.}]{Enhancement of acoustic spin pumping by acoustic distributed Bragg reflector cavity}

\author{Yunyoung Hwang}
 \affiliation{ 
Institute for Solid State Physics, University of Tokyo, Kashiwa, 277-8581, Japan
}

\author{Jorge Puebla}
\email{jorgeluis.pueblanunez@riken.jp}
\affiliation{ 
CEMS, RIKEN, 2-1, Hirosawa, Wako, 351-0198, Japan
}

\author{Mingran Xu}
 \affiliation{ 
Institute for Solid State Physics, University of Tokyo, Kashiwa, 277-8581, Japan
}

\author{Aurelien Lagarrigue}
\affiliation{ 
CEMS, RIKEN, 2-1, Hirosawa, Wako, 351-0198, Japan
}

\author{Kouta Kondou}
\affiliation{ 
CEMS, RIKEN, 2-1, Hirosawa, Wako, 351-0198, Japan
}

\author{Yoshichika Otani}
\email{yotani@issp.u-tokyo.ac.jp}
 \affiliation{ 
Institute for Solid State Physics, University of Tokyo, Kashiwa, 277-8581, Japan
}
\affiliation{ 
CEMS, RIKEN, 2-1, Hirosawa, Wako, 351-0198, Japan
}

\date{\today}

\begin{abstract}
Surface acoustic waves (SAWs) in the GHz frequency range can inject spin currents dynamically into adjacent nonmagnetic layers via spin pumping effect associated with ferromagnetic resonance. Here, we demonstrate an enhancement of acoustic ferromagnetic resonance and spin current generation by a pair of SAW reflector gratings, which form an acoustic analogue of the distributed Bragg reflector cavity. In the experiment, we confirmed 2.04 $\pm$ 0.02 times larger SAW power absorption in a device with cavity than in case of no acoustic cavity. We confirmed up to 2.96 $\pm$ 0.02 times larger spin current generation by measuring electric voltages generated by the inverse Edelstein effect (IEE) at the interface between Cu and Bi$_2$O$_3$. The results suggest that acoustic cavities would be useful to enhance the conversion efficiency in SAW driven coupled magnon-phonon dynamics.
\end{abstract}

\maketitle

The spin current, which is the flow of spins in a solid, it holds the promise of enabling efficient magnetic memories and computing devices in spintronics \cite{Wolf2001}, as well as acting as mediator in the interconversion between different physical entities.\cite{Otani2017} Therefore, it is desirable to find new routes to enhance the generation of spin current, which is commonly approach by appropriate selection of materials with engineered spin hall effect coefficients\cite{Nimi2015}. Here, we demonstrated the generation and enhancement of spin current by coupling surface acoustic waves with a ferromagnetic layer in the presence of an acoustic cavity. 

Surface acoustic waves (SAWs) in the GHz frequency range passing through a ferromagnetic layer can excite magnon-phonon dynamics, i.e., precessional magnetization motion mediated by the magnetoelastic effect.\cite{Ganguly1976,Feng1982,Wiegert1987}
This process is known as acoustic ferromagnetic resonance (A-FMR).\cite{Weiler2011,Dreher2012,Xu2018a,Puebla2020}
The A-FMR driven by SAWs can generate spin currents diffusing into adjacent non-magnetic metal layers via the spin pumping effect.\cite{Tserkovnyak2002a}
This coupled magnon-phonon dynamics can thus be used as a spin current generation method\cite{Weiler2012} named as acoustic spin pumping (ASP).\cite{Uchida2011,Uchida2012}
The generated spin currents can be detected by the inverse spin Hall effect (ISHE),\cite{Weiler2012,Saitoh2006a} or inverse Edelstein effect (IEE).\cite{Xu2018a,Sanchez2013}

In SAW driven A-FMR devices, interdigital transducers (IDT) are used for the generation and detection of SAWs.\cite{Feng1982,Wiegert1987,Weiler2011,Dreher2012,Xu2018a,Puebla2020,Weiler2012,Camara2019}
Since the SAWs propagate on both sides of the IDT, at least half of the phonon energy is lost. To reduce the loss and enhance the spin current generation via ASP, we employ acoustic cavity structures. An acoustic cavity (resonator) consists of a pair of acoustic wave reflector gratings, analogous to the distributed Bragg reflector for light.\cite{Bell1976}
In this letter, we demonstrate the enhancement of A-FMR and ASP in the presence of acoustic cavities.

\begin{figure}
\includegraphics[width=0.46\textwidth]{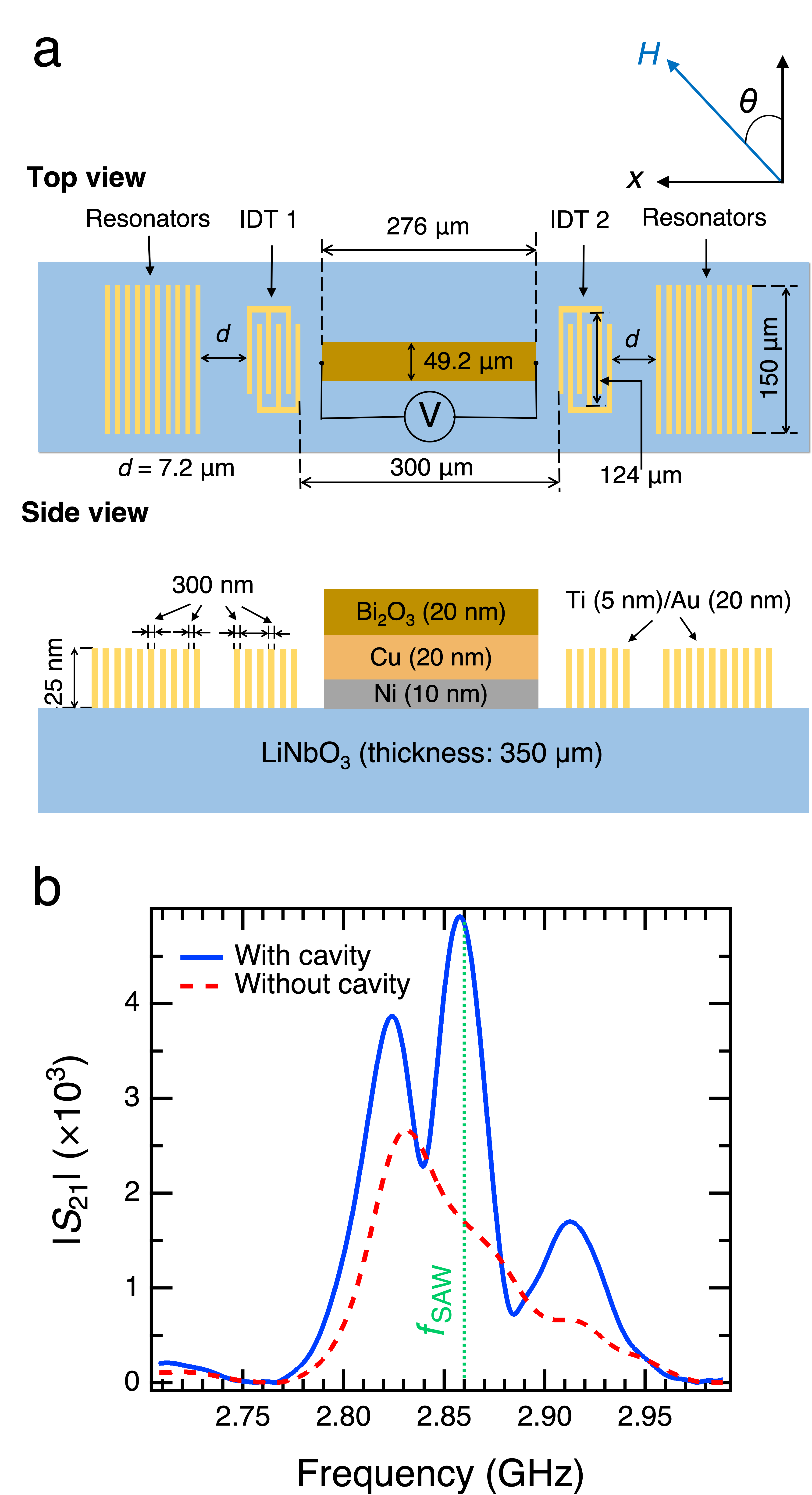}
\caption{\label{fig:Structure} Schematics of the ASP device structure with reflectors. Each IDT consists of two comb-shaped arrays of 50 metallic stripes, and each resonator comprises 200 reflector metallic stripes. The width and the gap of the metallic stripes of IDT and reflector are the same, 300 nm. SAWs are generated by applying RF voltage on IDT 1 or IDT 2. Generated SAWs propagate in $x$ and $-x$ direction and are reflected by the reflectors. (b) Measured SAW transmission in the frequency domain using a vector network analyzer (VNA). The measurements of the sample with an acoustic cavity (solid blue curve) and without an acoustic cavity (red dash line) are both displayed. The structures of the devices are the same except for the presence or not presence of reflectors.}
\end{figure}

For the generation of SAWs, we used a LiNbO$_3$ substrate on which we fabricated IDTs and reflector gratings using electron beam lithography. The dimensions of metallic electrodes in all IDTs are 300 nm in width, 25 nm in thickness, and the distance in-between metallic electrodes in all IDTs is 300 nm. Reflector gratings comprise of metallic wires with the same width and the distance between the closest gratings. After lithography, we deposited Ti (5 nm) and Au (20 nm) using thermal evaporation for IDTs and reflectors. The period of the IDT gratings gives the wavelength of the SAW $\lambda{}_{\mathrm{SAW}} = 1.2$ $\mu$m.\cite{Bell1976}
With the velocity of SAW on LiNbO$_3$ $v_{\mathrm{SAW}} = 3440$ m/s,\cite{Dreher2012} the resonance frequency of SAW $f_{\mathrm{SAW}}$ can be calculated as $f_{\mathrm{SAW}} = v_{\mathrm{SAW}}/\lambda{}_{\mathrm{SAW}} = 2.86$ GHz.\cite{Dreher2012,Puebla2020}
The distance between an IDT and reflector gratings $d$ is set to $d = 7.2$ $\mu$m, which is integer multiple of $\lambda{}_{\mathrm{SAW}}$. We fabricated a Ni (10 nm)/Cu (20 nm)/Bi$_2$O$_3$ (20 nm) trilayer stripe of $49.2 \times 276$ $\mu$m$^2$ by using electron beam evaporation and photolithography. The SAW excites A-FMR, i.e., the precessional magnetization dynamics in a bottom Ni layer, which causes a diffusive spin current pumped into the Cu/Bi$_2$O$_3$ Rashba interface. Consequently, the generated spin current is converted to charge current at the interface via IEE.\cite{Karube2016,Tsai2018}
The schematic illustration of our device is shown in FIG. \ref{fig:Structure}(a).

\begin{figure}
\includegraphics[width=0.45\textwidth]{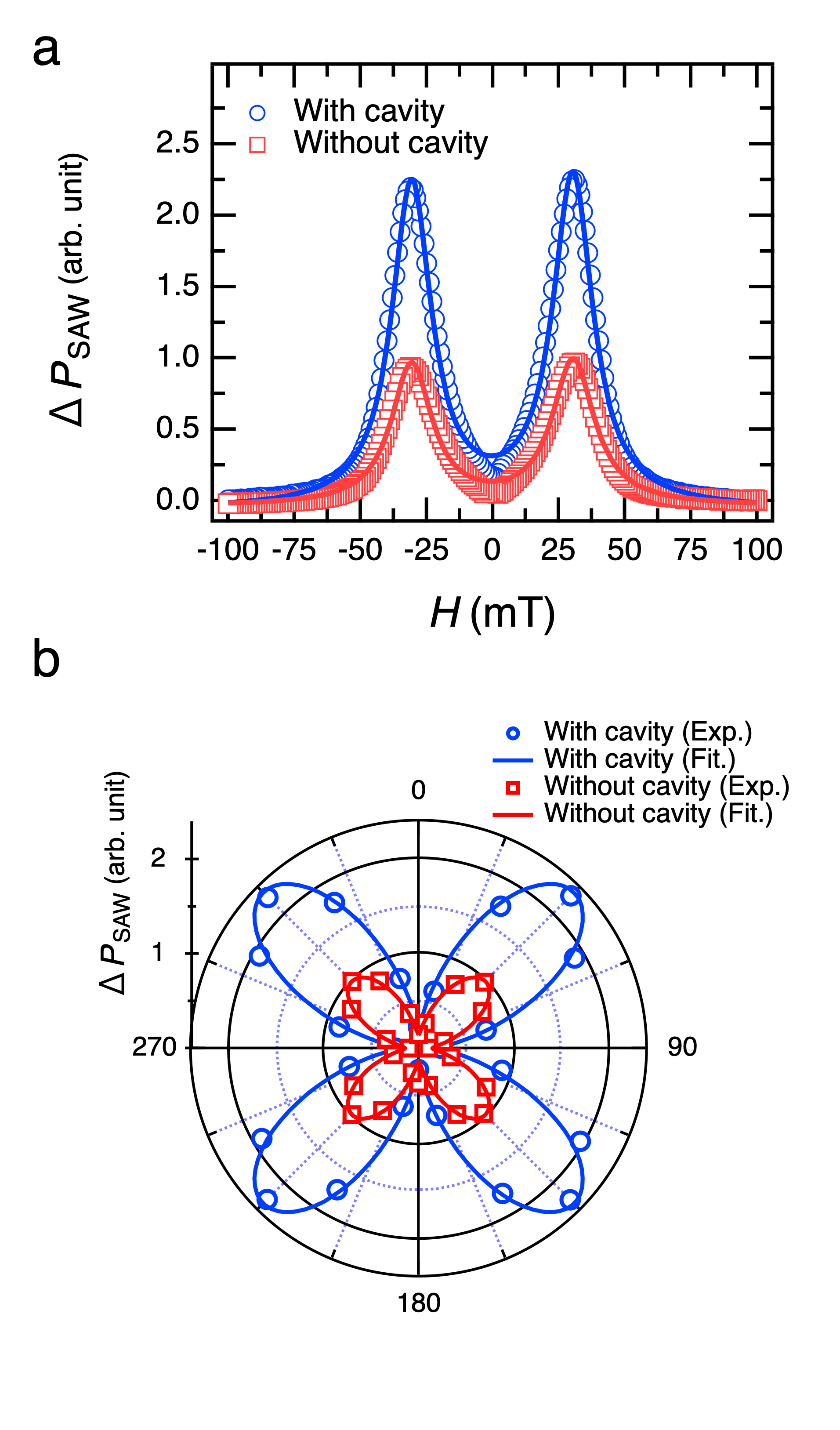}
\caption{\label{fig:abs} (a) Power absorption $\Delta{}P_{\mathrm{SAW}}$ of SAW at resonance condition for FMR driving of a Ni layer. (a) $\Delta{}P_{\mathrm{SAW}}$ at in-plane magnetic field $\theta = 45^{\circ}$ at the SAW resonance peak of samples with (blue circles) and without (red squares) acoustic cavity. 10 mW of input RF power is used. Solid curves exhibit Lorentzian fitting curves from the experimental data. (b) In-plane magnetic field angle $\theta$ dependence of absorbed SAW power of samples with (blue circles) and without (red squares) acoustic cavity. $\Delta{}P_{\mathrm{SAW}}$ is normalized with a $|S_{21}|$ value at $H = 100$ mT. Solid curves exhibit the fitting curves with Eq. (\ref{eq:abs}).}
\end{figure}

\begin{figure*}
\includegraphics[width=0.8\textwidth]{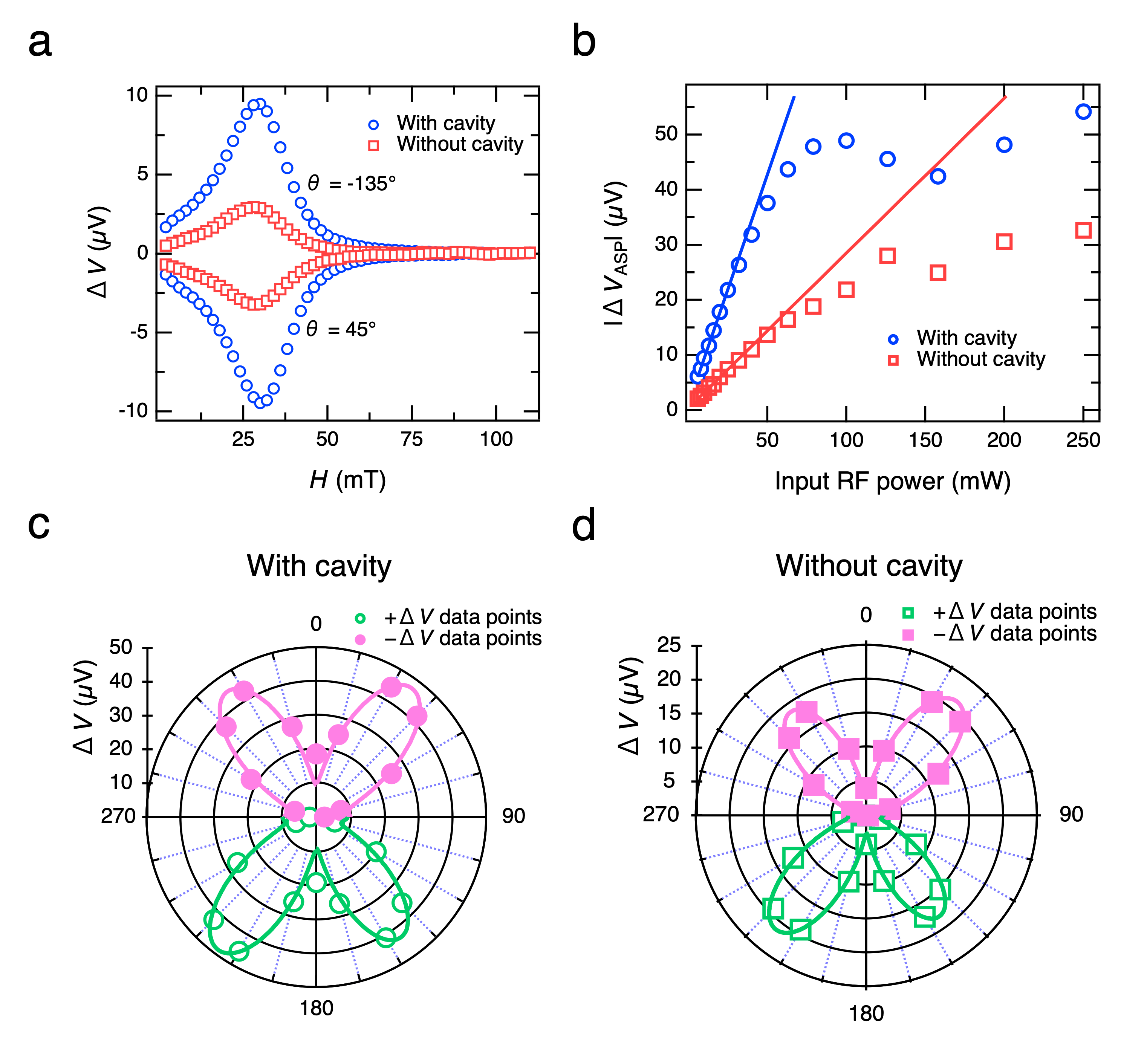}
\caption{\label{fig:IEE} (a) Detected electric voltage when a 10 mW of RF power is applied to IDT 1 and an external in-plane magnetic field $\theta = 45^{\circ}$ and $-135^{\circ}$ is applied. Data is taken from samples with acoustic cavity (blue circles), and without (red squares) acoustic cavity. The detected voltage is normalized with the voltage value at $H = 100$ mT ($\Delta{}V = V-V_{100 \mathrm{mT}}$). (b) Input RF power of IDT 1 dependence of $|\Delta{}V_{\mathrm{ASP}}|$. $\theta = 45^{\circ}$ of external magnetic field is applied to samples with (blue circles) and without (red squares) acoustic cavity. Solid (dash) lines exhibit fitting lines of the experimental data in linear behavior at low (high) power range. (c-d) In-plane magnetic field angle $\theta$ dependence of IEE voltage of samples (c) with and (d) without acoustic cavity. $\Delta{}V$ is normalized maximum value with a value at $H = 100$ mT. Solid curves exhibit fitting curves with Eq. (\ref{eq:dV}). Open green (solid pink) symbols exhibit positive (negative) $\Delta{}V$ points.}
\end{figure*}

\begin{figure}
\includegraphics[width=0.45\textwidth]{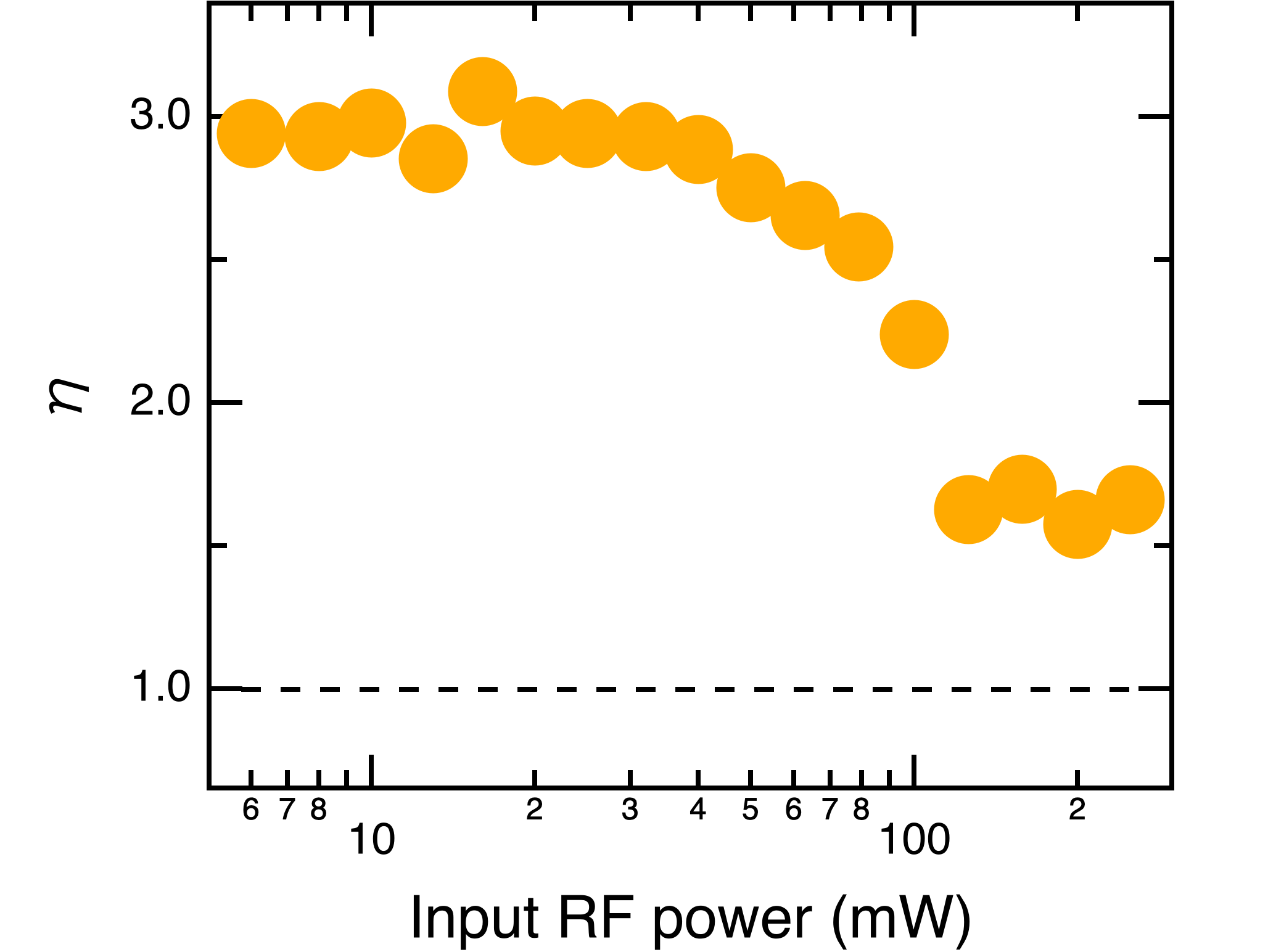}
\caption{\label{fig:ratio} Input RF power dependence of the ratio of $\Delta{}V_{\mathrm{ASP}}$ of samples with and without acoustic cavity ($\eta{}$) shown in FIG. \ref{fig:IEE}(b).}
\end{figure}

First, to confirm an enhancement of SAW amplitude using the acoustic cavity, we measure the SAW transmission using a vector network analyzer (VNA). The scattering parameter represents the SAW transmission $|S_{21}|$ defined as the radiofrequency (RF) power ratio of the electromagnetic wave detected and applied at IDT 1 and IDT 2, respectively.
The $|S_{21}|$ measurement result is shown in FIG. \ref{fig:Structure}(b).
The transmission spectra of our device with acoustic cavity shows 3 main peaks, being the central peak in close agreement with the estimated frequency in our device design ($f_{\mathrm{SAW}}=2.86$ GHz). The side peaks are likely coming from additional mechanical modes.\cite{Manenti2017} 
We find an enhancement factor of $1.7 \pm 0.2$ in the SAW power when using acoustic cavity. With a simulation of SAW confinement, we observe 2.2 times enhanced SAW amplitude (see Supplemental Material).

By applying an external magnetic field, SAWs passing Ni film induce A-FMR. When the A-FMR occurs, SAW power is attenuated due to energy conservation, thus the SAW power absorption is proportional to the induced A-FMR intensity $\Delta{}P_{\mathrm{SAW}} \propto [\mu_0 h_{\mathrm{RF}}]^2$, and\cite{Weiler2011,Dreher2012,Xu2018a,Puebla2020}
\begin{equation}\label{eq:abs}
  \left[ \mu_0 h_{\mathrm{RF}}\right]^2 = \left\{ b_1 \varepsilon_{xx}\sin{\theta}\cos{\theta} + 2 b_2 \varepsilon_{xz} \sin{\theta} \right\}^2,
\end{equation}
where $b_{1(2)}$ is the magnetoelastic coupling constant, $\varepsilon_{xx(xz)}$ represents the longitudinal (shear) strain induced by SAW \cite{Dreher2012,Xu2018a}.
The measured SAW power absorption in FIG. \ref{fig:abs}(a) shows the A-FMR signal fitted with a Lorentzian curve. Since the in-plane magnetic field angle dependence of SAW power absorption is well fitted with the Eq. (\ref{eq:abs}), as shown in the FIG. 2(b), we confirmed that the SAW power absorption is due to the A-FMR. By comparing amplitudes of the SAW power absorption from the fitting, we confirmed an enhancement of $2.04 \pm 0.02$ times A-FMR on the sample with acoustic cavity.

The A-FMR in the Ni layer generates spin current into the Cu layer by the ASP. The generated spin current is converted to charge current at the interface between Cu and Bi$_2$O$_3$ via IEE. We detect the generated electric voltage via IEE at the maximum A-FMR field angles $\theta = 45^{\circ}$ and $-135^{\circ}$, as shown in FIG. \ref{fig:IEE}(a). Since the ISHE is negligible in our device, the spin pumping signal is mainly caused by the IEE.\cite{Xu2018a}
$\Delta{}V$ is mainly from the sum of the voltage coming from IEE and anisotropic magnetoresistance (AMR). Both components can be separated by fitting because the IEE signal has symmetric Lorentzian shape and the AMR signal has antisymmetric Lorentzian shape.\cite{Mosendz2010}
We separate the component due to the ASP from symmetric and antisymmetric Lorentzian fitting. We define $\Delta{}V_{\mathrm{ASP}}$ as the amplitude of the symmetric Lorentzian fitting of $\Delta{}V$. The detected electric voltage from the ASP is caused by charge current, derived from an electric field $\mathrm{\mathbf{E}}$ induced by the IEE, which is proportional to the flow direction of the spin current density $\mathrm{\mathbf{J}}_s$ and its spin polarization $\bm{\sigma}_s$; $\mathrm{\mathbf{E}} \propto \mathrm{\mathbf{J}}_s \times \bm{\sigma}_s$.\cite{Karube2016} The converted charge current density $\mathrm{\mathbf{J}}_c$ is described as $\mathrm{\mathbf{J}}_c = \lambda_{\mathrm{IEE}} \mathrm{\mathbf{J}}_s$, where $\lambda_{\mathrm{IEE}}$ is the IEE length.\cite{Weiler2012}
As shown in FIG. \ref{fig:Structure}(a), since $\Delta{}V_{\mathrm{ASP}}$ is detected along the SAW propagation direction, it is described as $\Delta{}V_{\mathrm{ASP}} = \mathrm{\mathbf{J}}_{c}wR \sin{\theta}$, where $w$, $R$ are the width and the electric resistance of the Ni/Cu/Bi$_2$O$_3$ trilayer, respectively. Therefore, the magnitude of the generated spin current $J_s$ via coupled magnon-phonon dynamics in our samples can be estimated by
\begin{equation}\label{eq:spinCurrent}
  J_s = \frac{\Delta{}V_{\mathrm{ASP}}}{\lambda_{\mathrm{IEE}} wR\sin{\theta}}.
\end{equation}
Since the Ni/Cu/Bi$_2$O$_3$ trilayers of all samples are fabricated at the same time and the applied external magnetic field angle is the same in FIG. \ref{fig:IEE}(a), the parameters $\lambda_{\mathrm{IEE}}$, $w$, $R$, and $\theta$ are the same. Therefore, $\Delta{}V_{\mathrm{ASP}} \propto J_s$. While using the same measurement setting with FIG. \ref{fig:IEE}(a), we measured input RF power dependence of $\Delta{}V_{\mathrm{ASP}}$. The result is shown in FIG. \ref{fig:IEE}(b). With the maximum value of $\Delta{}V_{\mathrm{ASP}}$ in FIG. \ref{fig:IEE}(b), $\lambda_{\mathrm{IEE}} = -0.17$ nm,\cite{Tsai2018} $w = 49.2$ $\mu$m, $R = 70$ $\Omega$, and $\theta = 45^{\circ}$, we obtain a generated spin current $J_s = 1.3 \times 10^8$ A/m$^2$ via ASP.

We measure the in-plane magnetic field angle dependence of $\Delta{}V$. The measurement result is shown in FIG. \ref{fig:IEE}(c-d). We fit our data using the following equation\cite{Weiler2011}:
\begin{equation}\label{eq:dV}
    \Delta{}V_{\mathrm{ASP}} = \left[\mu_0 h_{\mathrm{RF}}\right]^2 \left[ C_1 \sin{\theta} + C_2 \cos{\theta} \right].
\end{equation}
As shown in FIG. \ref{fig:IEE}(c-d), both with and without acoustic cavity samples have similar fourfold symmetry. As discussed in ref. \onlinecite{Xu2018a}, the right and left side asymmetry in FIG. \ref{fig:IEE}(c-d) comes from the anisotropic distribution of charge potential in our measurement geometry.

We define the ratio of $\Delta{}V_{\mathrm{ASP}}$ of the sample with acoustic cavity and without acoustic cavity as $\eta$. FIG. 4 summarizes the input RF power dependence of $\eta$ derived from the data in FIG. 3(b), indicating that $\Delta{}V_{\mathrm{ASP}}$ varies linearly with the input power in the low-power range (< 25 mW). We estimate the enhancement factor as the ratio of the slopes of the fitting lines in FIG. 3(b). The enhancement factor of $\Delta{}V_{\mathrm{ASP}}$ is $2.96 \pm 0.02$ at the low-power range, and $1.6 \pm 0.7$ at the high-power range. However, the power dependence of $\Delta{}V_{\mathrm{ASP}}$ has non-linear behaviour in the range from 25 mW to 126 mW. This non-linearity in the power dependence is not fully understood. Since this behavior is similar to the case of FMR experiments at the high input RF power,\cite{Castel2012,Jungfleisch2015} we assume it is from the saturation of magnetic precessional cone angle\cite{Rana2017} or multi-magnon scattering.\cite{Jungfleisch2011,Chumak2012}
We observe a more significant enhancement factor of ASP in the low-power range than the enhancement of A-FMR. In contrast, we find a similar enhancement factor of ASP in the high-power range with that enhancement of A-FMR. The enhancement factor in the high-power range is well described by the multi-magnon scattering in ref. \onlinecite{Tashiro2015}. However, as far as the author knowledge goes the origin of the higher enhancement factor in the low-power range has not been observed and further understanding is required. 

Since A-FMR mainly occurs as spin-wave resonance (SWR),\cite{Li2017} we calculate the saturation magnetization $M_s = 0.28$ T with the resonant peak value derived from the measurement data of A-FMR and IEE (see the detail of the calculation in Supplemental Material). This value is 2 times smaller than the common $M_s$ value of bulk Ni. The origin of the suppressed $M_s$ value is missing, however, the similar $M_s$ value of 10 nm thickness of Ni has been reported.\cite{Wedler1977}

In summary, we have demonstrated the enhancement of A-FMR and spin current generation by using acoustic cavities. Enhancement of $2.04 \pm 0.02$ times of A-FMR and enhancements of spin current generation from $1.6 \pm 0.7$ (at high input RF power) to $2.96 \pm 0.02$ (at low input RF power) times were achieved. All the measurements in the present study were carried out at room temperature. At lower temperatures, the SAW confinement can be strengthen by minimizing interaction with thermal phonons. Minimization of phonon energy losses by further engineering of the acoustic cavities, as well as minimization of magnon energy losses by appropriate selection of materials may lead to magnon-phonon studies in the strong coupling regime.

The data that support the findings of this study are available from the corresponding author upon reasonable request.

\nocite{*}
\bibliography{AcousticCavities}

\clearpage
\widetext

\section*{Supplemental Material}
\subsection*{S1. Simulation about surface acoustic wave (SAW) confinement using acoustic cavity}

We performed a simulation to check how much SAW amplitude is enhanced using acoustic cavity. The simulation has been performed with COMSOL MULTIPHYSICS software. We set input interdigital transducers (IDT) and output IDT on a LiNbO$_{3}$ substrate. Due to the limited memory, we set 30 $\mu$m of distance between 2 IDTs while we used 300 $\mu$m of distance between 2 IDTs in the experiment. We also set 8 and 100 Au stripes of an IDT and a resonator, respectively while the real sample has 25 and 200 stripes of an IDT and a resonator. The structure geometry used in this simulation is demonstrated in Fig. S1. We apply input ac voltage which has same frequency with SAW to the boundary between IDT stripes and the substrate. SAW velocity and wavelength are set to 3440 m/s and 1.2 $\mu$m, respectively. We set 7.2 $\mu$m of distance between and IDT and a resonator as described in Fig. S1. We have done time-dependent simulation within a time range 0 to 45 ns. The simulation result is shown in Fig. S2. From sinusoidal fitting of a time range 20 to 40 ns, we confirm 2.23 times enhanced SAW amplitude by using acoustic cavity.

\renewcommand{\thefigure}{S1} 
\begin{figure}[h]
\includegraphics{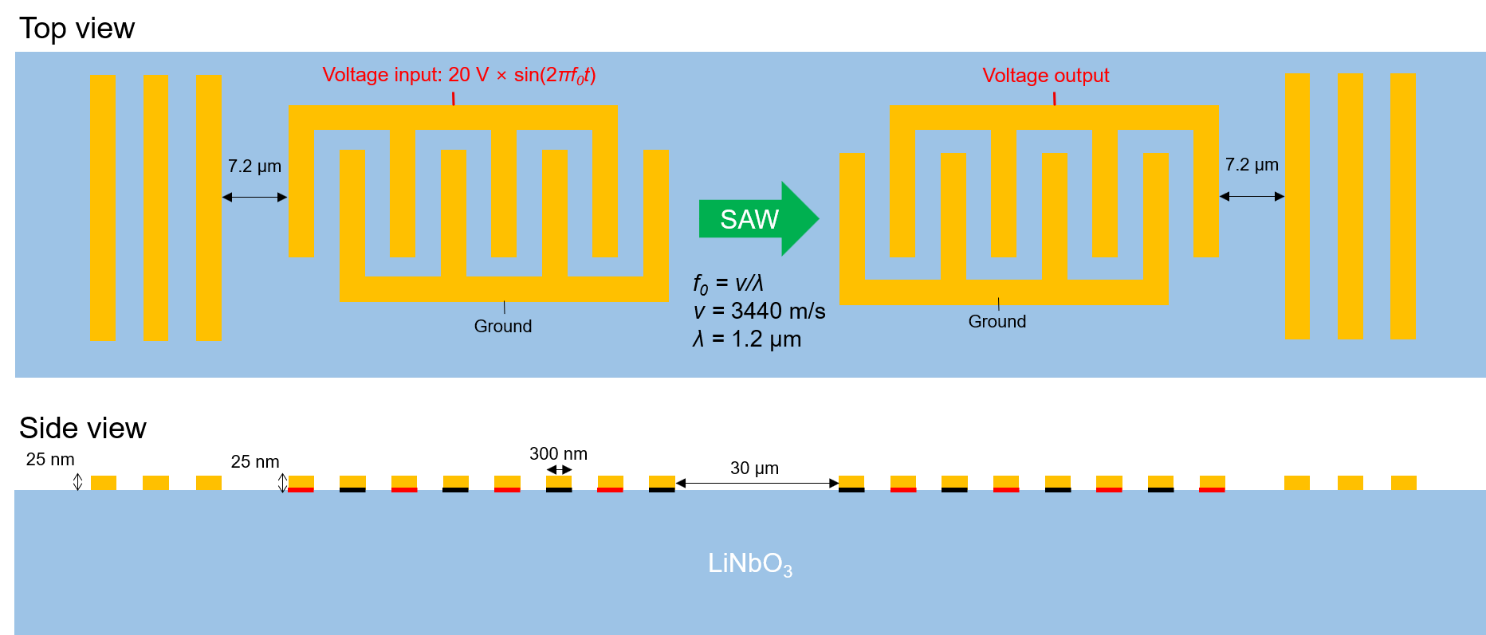}
\caption{Schematics of the structure used in this simulation. SAW velocity $v$, wavelength $\lambda$, frequency f$_0$ are described in the figure. The thickness of IDT and resonator stripes are both 25 nm. Each IDT has 8 Au stripes and each resonator has 100 Au stripes. The distance between 2 IDTs is 30 $\mu$m. The distance between IDT and a resonator is 7.2 $\mu$m. Input ac voltage which has same frequency with SAW applied the boundary between the input IDT and the substrate. A terminal for output voltage is set to the boundary between the output IDT and the substrate.}
\end{figure}

\renewcommand{\thefigure}{S2}
\begin{figure}[h]
\includegraphics{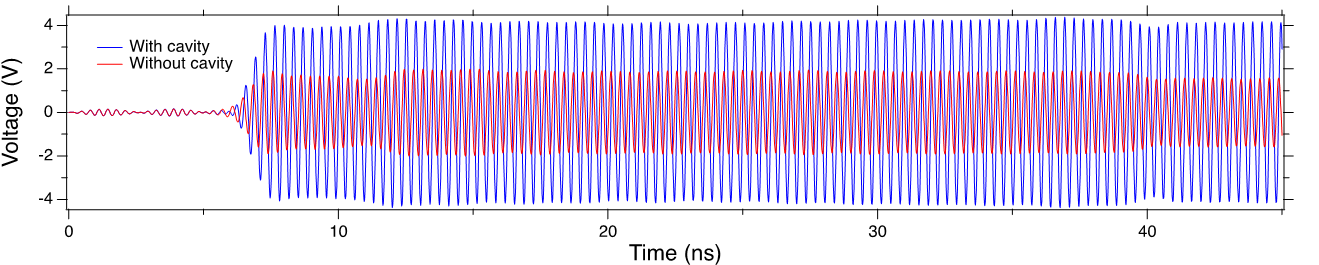}
\caption{Time-dependent simulation result. The result shows output ac voltage from the output IDT with (blue solid line) and without (red solid line) the presence of acoustic cavity.}
\end{figure}

\subsection*{S2. Acoustic ferromagnetic resonance (A-FMR) fitting data}

We show our A-FMR fitting here. Lorentzian fitting coefficients of samples are shown in Table S1. All measurements have done in $\theta$=45$^{\circ}$ of the field angle. We confirm 2.04 $\pm$ 0.02 times enhanced amplitude of dP using acoustic cavity.

\begin{table}[h!]
 \begin{center}
    \caption{A-FMR fitting data. Sample names are those described in Fig. 3}
    \label{tab:table1}
   \begin{tabular}{c|c|c|c} 
       \textbf{Sample} & \textbf{Amplitude of $\Delta$P$_{SAW}$ arb. units} & \textbf{Linewidth (FWHM)} & \textbf{Resonant peak (mT)}\\ 
      \hline
     Without cavity & 1.15 $\pm$ 0.02 & 19.44 $\pm$ 0.03 & 29.062 $\pm$ 0.005\\
      With cavity & 2.34 $\pm$ 0.01 & 19.5 $\pm$ 0.2 & 30.12 $\pm$ 0.05\\ 
    \end{tabular}
  \end{center}
\end{table}

\subsection*{S3. Inverse Edelstein effect (IEE) fitting method}

We show our measurement detail of IEE. We measure magnetic field H dependence of IEE voltage $\Delta V$ while we keep $\theta$=45$^{\circ}$ of the field angle and 10 mW of the input RF voltage to IDT 1. The fitting is done using the following equation:

\begin{equation}
\Delta V = \Delta V_{ASP} \frac{\Delta H^2}{(H-H_{res})^2+\Delta H^2}+\Delta V_{AMR} \frac{\Delta H (H-H_{res})}{(H-H_{res})^2+\Delta H^2},
\end{equation}

where $\Delta V_{ASP(AMR)}$, $\Delta H$ are the amplitude of the symmetric (anti-symmetric) Lorentzian curve and the line width (full width at half maximum, FWHM) of the Lorentzian peak.

\subsection*{S4. Spin wave resonance (SWR) calculation} 

We calculate the saturation magnetization M$_s$ of Ni in our Ni/Cu/Bi$_2$O$_3$ trilayer. of SWR which occurred by our SAW device. The relationship of the resonant field H$_{res}$ and the resonant frequency $f_0$ can be expressed as the following equations:

\begin{equation}
f_{0}=\frac{\gamma}{2\pi}\sqrt{(H_{res} + \mu_0 M_v)(H_{res} + \mu_0 M_z)}
\end{equation}

where

\[
  \begin{cases}
     M_v=\frac{Ak^2}{\mu _0 M_s}  + M_s \left( 1- \frac{1-e^{-kd}}{kd} \right) \sin ^2 \theta \\
    M_z=\frac{Ak^2}{\mu _0 M_s}  + M_s \frac{1-e^{-kd}}{kd}  
  \end{cases}
\]

where $\gamma$, $\mu_0$, $A$, $k$, $d$, $\theta$ are the gyromagnetic ratio, the vacuum permeability, exchange stiffness, wavenumber, the thickness of the ferromagnetic layer and the magnetic field angle, respectively. We employ $\gamma = 1.76 \times 10^{-11}$, $A = 1.05 \times 10^{-11}$ J/m, $k = 1/1.2$ $\mu m^{-1}$, $d = 10$ nm, $\theta = 45^{\circ}$, $H_{res} = 30$ mT. From the calculation of Eq. (2) with these parameters of Ni, we derived $M_s = 2.2 \times 10^5$ A/m.

\end{document}